\documentclass[10pt,showpacs,preprintnumbers,amsmath,amssymb]{revtex4}
\usepackage{units}
\usepackage{bbold,mathpazo}
\usepackage{graphicx}
\usepackage{dcolumn}
\usepackage{bm,bbm,mathtools,esvect}
\usepackage{slashed}

\begin{document}

\title{NON-ANTI-HERMITIAN QUATERNIONIC QUANTUM MECHANICS}
\vspace{2cm}
\author{SERGIO GIARDINO}
\email{giardino.sergio@unifesp.br}
\affiliation{\vspace{3mm} Institute of Science and Technology, Federal University of S\~ao Paulo\\
Avenida Cesare G. M. Lattes 1201, 12247-014 S\~ao Jos\'e dos Campos, SP, Brazil}

\begin{abstract}
\noindent 
The breakdown of Ehrenfest's theorem imposes serious limitations on quaternionic quantum mechanics (QQM). 
In order to determine the conditions in which the theorem is valid,  
we examined the conservation of the probability density, the expectation value and the classical limit 
for a non-anti-hermitian formulation of QQM. The results also indicated that the non-anti-hermitian quaternionic
theory is related to non-hermitian quantum mechanics, and thus the physical problems described with both of the theories should be related.
\end{abstract}

\maketitle
\tableofcontents
\section{Introduction\label{I}}
Currently, quaternionic quantum mechanics (QQM) is a theory of anti-hermitian operators \cite{Adler:1995qqm}, and 
thus the mathematical framework of QQM has been developed using hermitian formalism of 
complex quantum mechanics (CQM) as a reference frame. Anti-hermiticity is a way to preserve the probability 
density \cite{Adler:1995qqm}, and thus anti-hermiticity is believed as necessary to a consistent QQM. Nevertheless, a quaternionic 
non-anti-hermitian solution that preserves the proability density
has recently been obtained \cite{Giardino:2016xap}, and thus the question about the necessity
of the anti-hermitian assumption is posed. Furthermore, quaternionic definitions for the momentum operator,
for the probability current and for the expectation values are proposed in that article. These results points out that several 
CQM structures have no exact equivalent in QQM. Consequently, the physical interpretation of the theories may differ dramatically. 

In this study, we are interested in the specific discrepancy between
the classical limits of hermitian CQM and anti-hermitian QQM. The Ehrenfest theorem states that the expectation values of
position and linear momentum calculated through CQM obey a classical dynamics. The theorem thus states that quantum mechanics and
classical mechanics are somewhat related, so that quantum dynamics must have a classical limit. Furthermore, it sets forth the background for proposing that quantum phenomena may be 
generated by fluctuations of classical quantities. Accordingly, the Ehrenfest theorem is a basic concept that enables the formulation of
semi-classical quantum mechanics, with far-reaching consequences that are both conceptual and practical in nature. 
Consequently, the breakdown of the Ehrenfest theorem for anti-hermitian QQM \cite{Adler:1985uh,Adler:1988fs,Adler:1995qqm} proves that the physical
contents of hermitian CQM and anti-hermitian QQM are different, and thus the phenomena described by 
both  theories are probably different. 

We may conclude that QQM is either disconnected from classical mechanics, and thus QQM has no
classical limit, or that within the classical limit of QQM there is a generalized, and unknown, classical theory. 
In this study, we propose another point of view, in which QQM is not an anti-hermitian generalization of hermitian CQM. 
In fact, we propose 
a non-anti-hermitian QQM as a generalization for non-hermitian CQM \cite{Bender:2007nj,Moiseyev:2011nhq}. 
Using this simple assumption, we were able to 
ascertain that the breakdown of Ehrenfest for non-anti-hermitian Hamilton operators
is similar to the breakdown of the Ehrenfest theorem observed in non-hermitian CQM, and thus we expect that a 
link between non-hermitian
CQM and non-anti-hermitian QQM may be established physically as well as mathematically.
Furthermore, we shall see that the
Ehrenfest theorem may be verified in the particular case where hermitian operators are considered for QQM. 

On the other hand, if non-anti-hermitian QQM is somewhat related to non-hermitian CQM, we may have an important way
of testing QQM. We remember that non-hermitian CQM has deserved an remarkable interest from experimental physicists 
\cite{Makris:2008dyn,Guo:2009obs,Rueter:2010pts,Regensburger:2012phl}, and these techniques may test QQM as well.
Furthermore, there exist different theoretical proposals about non-hermitian CQM 
\cite{Scolarici:2003wu,Scolarici:2009zz,Solombrino:2002vk,Mostafazadeh:2001jk,Mathur:2010nhh,Mathur:2014rnh} and even for 
non-anti-hermitan QQM \cite{Blasi:2005alt,Blasi:2004qah}. In summary, the proposal of this article is reltated to important
trends in quantum mechanics which we hope will increase the interest in QQM and enable us to understand which
 kind of physical phenomena may be described with it, if any. 

This article is organized according to the two possible quaternionic wave equations that we 
consider, namely the left complex wave equation (LCWE) and the right complex wave equation (RCWE). In Section \ref{A} we 
define 
the LCWE and study its fundamental properties, namely the continuity equation for the probability density, the expectation
values and the Ehrenfest theorem. Furthermore, we study basic properties of hermitian Hamiltonians in QQM. In section
\ref{B} we repeat the results for RCWE and, in Section \ref{C}, we present our conclusions and future directions for 
research.

\section{\label{A} The left-complex wave equation}
Quaternions ($\mathbb{H}$) are generalized complex numbers with three anti-commuting complex units: $i,\;j$ and $k$ \cite{Rocha:2013qtt}. The complex
units satisfy 
\begin{equation}\label{A4}
ij=-ji=k\qquad \mbox{and}\qquad ijk=-1,
\end{equation}
and an arbitrary quaternionic number is written as
\begin{equation}
 q=x_0+x_1 i+x_2 j+ x_3 k,
\end{equation}
where $x_0,\,x_1,\,x_2$ and$ x_3$ are real. In symplectic notation, $q\in\mathbb{H}$ is written
\begin{equation}\label{a4}
 q=z +\zeta j\qquad\mbox{with}\qquad z,\,\zeta\in\mathbb{C}.
\end{equation}
Let us consider the quaternionic Schr\"odinger equation 
\begin{equation}\label{A1}
i\hbar\,\frac{\partial\Psi}{\partial t}= \mathcal{H}\Psi,
\end{equation}
where $\Psi$ and $\mathcal{H}$ are quaternionic. The left hand side of (\ref{A1}) admits two positions of the complex unit $i$, and hence we
call  (\ref{A1}) the left-complex wave equation (LCWE). In accordance with a previous study of the Aharonov-Bohm 
effect in QQM \cite{Giardino:2016xap}, we propose the quaternionic Hamiltonian operator
\begin{equation}\label{A2}
 \mathcal{H}=\frac{\hbar^2}{2m}i\Big(\bm\nabla -\bm Q\Big)\bm\cdot i\Big(\bm\nabla -\bm Q\Big)+V.
\end{equation}
$\bm Q$ is a pure imaginary quaternionic vector, and $V$ is a quaternionic
scalar potential. Using the symplectic notation we write
\begin{equation}\label{A3}
\bm Q=\bm\alpha\, i+\bm\beta\,j\qquad\mbox{and}\qquad V=V_0+V_1\,j,
\end{equation}
where $\bm\alpha$ is real and $\bm\beta,\;V_0$ and $V_1$ are complex. We notice that a quaternionic imaginary vector potental
has been introduced by Michael Atyiah for examining Yang-Mills instantons \cite{Atiyah:1979gym}, and this potential
has been firstly used in QQM considering the quaternionic Aharonov-Bohm effect \cite{Giardino:2016xap}.
 The quaternionic Hamiltonian
(\ref{A2}) is general and neither Hermiticity nor anti-Hermiticity are supposed.  Furthermore, we whish to examine the conservation
of probability in (\ref{A1}). If $\rho$ is the probability density, thus
\begin{equation}\label{AA1}
\frac{\partial}{\partial t}\int dx^3\rho=\int dx^3\left(\frac{\partial\Psi^*}{\partial t}\Psi+
\Psi^*\frac{\partial\Psi}{\partial t}\right)\qquad\mbox{where}\qquad\rho=\Psi^*\Psi.
\end{equation}
Furthermore, using (\ref{A1}) and (\ref{A2}), we obtain
\begin{equation}\label{AA2}
\Psi^* \frac{\partial\Psi}{\partial t}=\frac{\hbar}{2m}\left\{\bm{\nabla\cdot}\Big[\Psi^*i(\bm\nabla\Psi-\bm Q\Psi)\Big]
+\bm\nabla\Psi^*\bm\cdot i\bm Q\Psi-\Psi^*\bm Q i\bm{\cdot\nabla}\Psi-
\bm\nabla\Psi^*i\bm{\cdot\nabla}\Psi-\Psi^*\bm Q\cdot i\bm Q\Psi
\right\}-\frac{1}{\hbar}\,\Psi^*iV\Psi.
\end{equation}
Several terms of the right hand side of (\ref{AA2}) cancel out in (\ref{AA1}). Consequently, the continuity equation reads
\begin{equation}\label{A5}
\frac{\partial \rho}{\partial t}+ \bm\nabla\cdot \bm J=g,
\end{equation}
where the probability current $\bm J$, the gauge-invariant quaternionic linear 
momentum operator $\bm\Pi$, and the source $g$ are as follows
\begin{equation}\label{A6}
g=\Psi^*\frac{V^* i-i\, V}{\hbar}\Psi,\qquad
\bm J=\frac{1}{2m}\left[\Psi^*\big(\bm\Pi\Psi\big)+\big(\bm\Pi\Psi\big)^*\Psi\,\right],
\qquad\mbox{and}\qquad\bm\Pi\Psi=-i\,\hbar\big(\bm\nabla-\bm Q\big)\Psi.
\end{equation}
If the source is zero, there is neither source nor sink of probability, and this is an important consistency test. 
The source term $g$ of (\ref{A5}) is zero for real $V$, and  the correspondence between QQM and CQM is exact in this case.
 However, the conservation
of the probability density does not preclude the existence of other sources of discrepancy between QQM and CQM, as we
shall see for the Ehrenfest theorem. On the other hand, non-zero sources appear in non-Hermitian CQM 
\cite{Bender:2007nj,Moiseyev:2011nhq}, particularly when 
complex scalar potentials are admitted, and then we shall research quaternionic potentials taking the non-hermitian
complex case as our frame of reference.

Real $V$ potentials have been explored in non-anti-hermitian QQM with relative success 
\cite{Davies:1989zza,Davies:1992oqq,Ducati:2001qo,Nishi:2002qd,DeLeo:2005bs,Ducati:2007wp,Davies:1990pm,DeLeo:2013xfa,DeLeo:2015hza,Giardino:2015iia,Sobhani:2016qdp,Procopio:2016qqq}.
However, we point out that (\ref{A2}) is neither hermitian nor anti-hermitian.
This is an example that leads to the question of whether anti-hermiticity is really necessary for QQM.
In order to obtain a quaternionic quantum theory with a well-defined classical limit, we take inspiration from complex quantum
mechanics. From the probability current (\ref{A6}), we write the canonical momentum as
\begin{equation}\label{A7}
\langle \bm\Pi \rangle =m\int dx^3\bm J.
\end{equation}
Using the quaternionic probability current from (\ref{A6}), we propose the expectation value for an arbitrary quaternionic operator 
$\cal{O}$  to be
\begin{equation}\label{A8}
\langle\mathcal O\rangle= \frac{1}{2}\int dx^3\Big[\Psi^*\big(\mathcal{O}\Psi\big) +\Big(\Psi^*\big(\mathcal{O}\Psi\big) \Big)^* \Big].
\end{equation}
This expectation value is based on a quaternionic scalar product compatible to Fock and Hilbert spaces \cite{Giardino:2012ti}, and thus
a more rigorous study concerning the hermiticity of $\mathcal O$ can be conducted in future research. 
Definition (\ref{A8}) generalizes the expectation value of 
CQM for two reasons. Firstly, because the usual definition is recovered when $\mathcal O$ is 
hermitian and, secondly because (\ref{A8}) is real for every $\mathcal O$, regardless of its hermiticity.
Let us next ascertain whether QQM is well-defined within the classical limit when the expectation 
value (\ref{A8}) is supposed. The time-derivative of the position operator $\bm r$ gives
\begin{equation}\label{A9}
\frac{d \langle \rm r \rangle}{dt}=\frac{1}{m}\langle \bm\Pi\rangle-\frac{2}{\hbar}\big\langle iV\bm r\big\rangle
\end{equation}
This result is in agreement with hermitian CQM for real $V$ and is in agreement with non-hermitian
CQM for complex $V$, and thus we hypothesize 
 that non-anti-hermitian QQM may generalize non-hermitian CQM. By way of clarification, we notice that
\begin{equation}\label{A10}	
 2\big\langle iV\bm r\big\rangle=\big\langle (iV-V^*i)\bm r\big\rangle
\end{equation}
is identical to zero for real $V_0$, considering $V$ as defined  in (\ref{A3}). This means that 
$\langle \bm r \rangle$ obeys 
 classical dynamics, and consequently satisfies the Ehrenfest theorem for real $V_0$ and $|\bm\beta|=0$, a fact that is already known for anti-hermitian QQM \cite{Adler:1995qqm}. 
A real $V$ implies that $\langle \bm r \rangle$ is
dynamically classical and additionally that $\mathcal H$ is hermitian; this constitutes further evidence that anti-hermitian operators
may not be essential to QQM. (\ref{A9}) recovers the usual form of Ehrenfest theorem within the limit $\bm Q=\bm 0$,
where the usual linear momentum $\bm p$ replaces $\bm \Pi$.

Let us next consider whether the expectation 
value of the linear momentum operator also behaves like the position expectation value. Along the $x$ direction, we get
\begin{equation}\label{A11} 
\frac{d\langle p_x\rangle}{dt}=\int dx^3\left(\frac{\partial\Psi^*}{\partial x}V\Psi+\Psi^*V^*\frac{\partial\Psi}{\partial x}\right).
\end{equation}
For real $V$, the right hand side of (\ref{A11}) gives $\langle-\,\partial_x V\rangle$, in perfect agreement with hermitian CQM. 
Using expectation values, we obtain
\begin{equation}\label{A12} 
\frac{d\langle p_x\rangle}{dt}=2\left\langle - \frac{\partial V}{\partial x}\right\rangle+
2\left\langle -V\frac{\partial}{\partial x} \right\rangle.
\end{equation}
In this case, there is a perfect agreement  with CQM for
real and complex potentials. Additionally, the Ehrenfest theorem is verified for real V, and thus this case
 comprehends the quaternionic Aharonov-Bohm effect \cite{Giardino:2016xap} as well. This means that QQM does satisfy the
Ehrenfest theorem in the case of hermitian Hamiltonians, and the breakdown of the theorem for non-anti-hermitian
operators is in agreement with anti-hermitian CQM. This enables us to infer that the origin of the breakdown of the 
classicality of $\langle x \rangle$ and $\langle p_x \rangle$ have the same origin, namely the non-hermiticity of the
pure imaginary terms of the potential. Before considering the right complex wave equation case, let us consider 
several interesting properties of hermitian Hamiltonians in QQM. We notice that the 
time derivative of $\bm \Pi$ has not been presented because it seems too complicated and difficult to interpret.
However, once understood, this result will provide the quaternionic version of the quantum Lorentz force, 
and hence this point remains as an important direction for future research.

\subsection{ Hermitian Hamiltonian operators}

If $\mathcal H$ is hermitian, it may be interchanged with $i\hbar\partial_t$, regardless of the wave function. 
Using this fact, we use (\ref{A1}) and (\ref{A8}) to get the  identity
\begin{equation}\label{A14}
\big\langle \mathcal{HO}\big\rangle=\hbar\left\langle i\frac{\partial\mathcal{O}}{\partial t}\right\rangle-
\big\langle i\mathcal{O}i\mathcal{H}\big\rangle,
\end{equation}
where $\mathcal{HO}\Psi=i\hbar\partial_t(\mathcal{O}\Psi)$ has been used. Similar relations are obtained by replacing $\mathcal{O}$ with $i\mathcal{O}i$, $\mathcal{O}i$ and $i\mathcal{O}$. From them,
we eventually obtain
\begin{align}\nonumber
&\Big\langle\big[\mathcal{H},\,\mathcal{O}-i\mathcal{O}i\,\big]\Big\rangle=
\hbar\left\langle \frac{\partial}{\partial t}( \mathcal O i+ i\mathcal O )\right\rangle,
\qquad
\Big\langle\big\{\mathcal{H},\,\mathcal{O}+i\mathcal{O}i\,\big\}\Big\rangle=
-\hbar\left\langle \frac{\partial}{\partial t}(\mathcal O i - i\mathcal O )\right\rangle,
\\ \label{A15}
&\Big\langle\big[\mathcal{H},\,\mathcal{O}i+i\mathcal{O}\,\big]\Big\rangle=-
\hbar\left\langle \frac{\partial}{\partial t}(\mathcal O- i\mathcal O i)\right\rangle,
\qquad
\Big\langle\big\{\mathcal{H},\,\mathcal{O}i-i\mathcal{O}\,\big\}\Big\rangle=
\hbar\left\langle \frac{\partial}{\partial t}(\mathcal O + i\mathcal O i)\right\rangle,
\end{align}
where the square brackets denote commutation relations and the curly brackets denote anti-commutation relations.
The set of relations (\ref{A15}) assures that the quaternionic solutions of (\ref{A1}) are stationary states. In other words, 
we have the Schr\"odinger picture, where wave functions are time-dependent and the operators are time-independent. 
At this point, it is natural do discuss the time evolution for the expectation values. Assuming (\ref{A1}) and (\ref{A8}), 
we get the identities
\begin{align}
& \frac{d}{dt}\Big\langle \mathcal O - i\mathcal O i\Big\rangle
=\left\langle\frac{\partial}{\partial t}(\mathcal O - i\mathcal O i)\right\rangle
+\frac{1}{\hbar}\left\langle \Big[\mathcal H,\,\mathcal O i+i\mathcal O\Big]\right\rangle+
\frac{\partial}{\partial t}\Big\langle \mathcal O - i\mathcal O i\Big\rangle,\label{A16}
\\
& \frac{d}{dt}\Big\langle \mathcal O i + i\mathcal O \Big\rangle
=\left\langle\frac{\partial}{\partial t}(\mathcal O i +i\mathcal O )\right\rangle
-\frac{1}{\hbar}\left\langle \Big[\mathcal H,\,\mathcal O -i\mathcal O i\Big]\right\rangle+
\frac{\partial}{\partial t}\Big\langle \mathcal O i+ i\mathcal O \Big\rangle,\label{A17}
\\
& \frac{d}{dt}\Big\langle \mathcal O + i\mathcal O i\Big\rangle
=\left\langle\frac{\partial}{\partial t}(\mathcal O + i\mathcal O i)\right\rangle
-\frac{1}{\hbar}\left\langle \Big\{\mathcal H,\,\mathcal O i-i\mathcal O\Big\}\right\rangle+
\frac{\partial}{\partial t}\Big\langle \mathcal O + i\mathcal O i\Big\rangle,\label{A18}
\\
& \frac{d}{dt}\Big\langle \mathcal O i-i\mathcal O \Big\rangle
=\left\langle\frac{\partial}{\partial t}(\mathcal O i - i\mathcal O )\right\rangle
+\frac{1}{\hbar}\left\langle \Big\{\mathcal H,\,\mathcal O +i\mathcal O i\Big\}\right\rangle+
\frac{\partial}{\partial t}\Big\langle \mathcal O i- i\mathcal O \Big\rangle.\label{A19}
\end{align}
If
$\mathcal O$ and $\Psi$ are complex, (\ref{A16}) and (\ref{A17}) recover the usual CQM relation, while (\ref{A18}) and 
(\ref{A19}) become trivial and the last term of the right hand side of each (\ref{A16}-\ref{A19}) disappears. This fact
enables us to interpret that, in CQM, if (\ref{A15}) is valid then we will have stationary states. 

Conversely, using (\ref{A15}) 
in (\ref{A16}-\ref{A19}) we calculate that the total time-derivatives are not identical to zero. Thus, the expectation 
values are not necessarily independent of time, nor are the wave functions stationary states. In order to obtain a quaternionic
Schr\"odinger picture for LCWE we need the additional set of constraints, namely
\begin{equation}\label{A20}
\frac{\partial}{\partial t}\Big\langle \mathcal O - i\mathcal O i\Big\rangle=
\frac{\partial}{\partial t}\Big\langle \mathcal O i+ i\mathcal O \Big\rangle=
\frac{\partial}{\partial t}\Big\langle \mathcal O + i\mathcal O i\Big\rangle=
\frac{\partial}{\partial t}\Big\langle \mathcal O i-i\mathcal O \Big\rangle=0.
\end{equation}
Now, if (\ref{A15}) and (\ref{A20}) are valid, then we have stationary states and the quantum quaternionic states may be considered to
be framed in the Schr\"odinger picture. A Heisenberg picture and the Virial theorem are also valid for hermitian Hamiltonian 
operators in the same fashion as in CQM, and thus QQM and CQM are perfectly compatible for hermitian Hamiltonians.

\section{\label{B} The right complex wave equation}
In this section, we explore solutions of the quaternionic Schr\"odinger equation 
\begin{equation}\label{B1}
\hbar\,\partial_t\Psi i=\mathcal{H}\Psi,
\end{equation}
that we call the right complex wave function (RCWF). In this case, we have
\begin{equation}\label{BB22}
 \mathcal{H}=-\frac{\hbar^2}{2m}\Big(\bm\nabla -\bm Q\Big)^2+V.
\end{equation}
Following the LCWE case, we will study the continuity equation. Using (\ref{AA1}), (\ref{B1}) and (\ref{BB22})
\begin{equation}\label{BB2}
 \frac{\partial\Psi}{\partial t}\Psi^*=\frac{\hbar}{2m}\left\{\bm{\nabla\cdot}\Big[(\bm\nabla\Psi-\bm Q\Psi)i\,\Psi^*\Big]
+\bm{Q\cdot}\Big(\Psi\,i\bm\nabla\Psi^*-\bm\nabla\Psi i\,\Psi^*\Big)-\bm\nabla\Psi\bm\cdot (i\bm\nabla\Psi^*)-|\bm Q|^2\Psi i\,\Psi^*
\right\}-\frac{1}{\hbar}V\,\Psi i\,\Psi^*.
\end{equation}
Because $\;\Psi\,i\bm\nabla\Psi^*-\bm\nabla\Psi i\,\Psi^*\;$ is real and $\bm Q$ is pure imaginary, several terms cancel out in (\ref{AA1}). 
Thus, we obtain a continuity equation (\ref{A5}) where the source, the probability current and the linear momentum are such that
\begin{equation}\label{B2}
g=\frac{1}{\hbar}\Big(\Psi i\,\Psi^* V^*-V\Psi i\Psi^* \Big),\;\;\;\;
\bm J=\frac{1}{2m}\Big[(\bm\Pi\Psi)\Psi^*+\Psi(\bm\Pi\Psi\big)^* \Big]\;\;\;\;\mbox{and}\qquad
\bm\Pi\Psi=-\hbar\big(\bm\nabla-\bm Q\big)\Psi i.
\end{equation}
Hence, we propose the expectation value (\ref{A7}) as the expectation value 
\begin{equation}
\langle\mathcal O\rangle= \frac{1}{2}\int dx^3\Big[\big(\mathcal{O}\Psi\big)\Psi^* +\Big(\big(\mathcal{O}\Psi\big)\Psi^* \Big)^* \Big].
\end{equation}
We can accordingly study the Ehrenfest theorem, so that
\begin{equation}\label{B4}
\frac{d \langle \bm r \rangle}{dt}=\frac{\langle \bm \Pi\rangle}{m}-\frac{2}{\hbar}\big\langle (V\bm r|i)\big\rangle,
\end{equation}
where we define the notation
\begin{equation}\label{B5}
 (\mathcal O|i)\Psi=\mathcal O\,\Psi i.
\end{equation}
The second term on the right hand side of (\ref{B4}) are zero for real $V$,
and the dynamics of $\langle x \rangle$ is classical for hermitian Hamiltonians as well. If we calculate the expectation value for the linear
 momentum along the $x$ direction we obtain
\begin{equation}\label{B6} 
\frac{d\langle p_x\rangle}{dt}=\int dx^3\left(V\Psi\frac{\partial\Psi^*}{\partial x}+\frac{\partial\Psi}{\partial x}\Psi^*V^*\right).
\end{equation}
which recovers the CQM result for real $V$, as expected. (\ref{B6}) implies (\ref{A12}),
and thus Ehrenfest's theorem is valid for the RCWE dynamics as well.

\subsection{Hermitian Hamiltonian operators}
A hermitian Hamiltonian enables us to obtain
\begin{equation}\label{B7}
 \Big\langle \big[\mathcal H,\,\mathcal O\big]\Big\rangle=\hbar\left\langle\left(\frac{\partial}{\partial t}\mathcal O\Big|i\right)\right\rangle,
\end{equation}
and we observe that there are important differences compared to the LCWF: there is only one equation and there 
are no
anti-commutation relations. We expect that the physical content of the left complex case is different from the
right complex case, but the actual differences will only be ascertained after explicit solutions have been found.  Finally, we get
\begin{equation}
\frac{d}{dt}\Big\langle \big(\mathcal{O}|i\big) \Big\rangle
=\left\langle\frac{\partial}{\partial t}\big(\mathcal O|i\big) \right\rangle
+\frac{1}{\hbar}\left\langle \Big[\mathcal O,\,\mathcal H \Big]\right\rangle
+\frac{\partial\langle(\mathcal O|i)\rangle}{\partial t}.\label{B8}
\end{equation}
As in the LCWE, stationary states are obtained if a set of constraints include (\ref{B7}) and 
\begin{equation}
\frac{\partial\langle\mathcal O\rangle}{\partial t}=0. 
\end{equation}
Hence we have a consistent QQM for the RCWE, which contains a wave equation, a continuity equation and a classical limit.
In future research, we will develop explicit solutions to illustrate and build models where some physical phenomena can be researched. 
However, the important point of the formal consistent has been established throughout this study.

\section{\label{C} Conclusion}
In this article, we have proposed an alternative formulation for quaternionic quantum mechanics that has
enabled us to explain the breakdown of the Ehrenfest theorem 
observed in the anti-hermitian formulation of QQM. This formulation of QQM encompasses  
quaternionic Hamiltonians, and neither hermiticity
nor anti-hermiticity are supposed. In spite of this, we were able to define a theory with real-value 
expectation values. The question about the necessity of the anti-hermitian assumption in QQM has arisen in
a non-anti-hermitian solution to the quaternionic Aharonov-Bohm effect \cite{Giardino:2016xap}, and the present article has
given a formal expression to a non-anti-hermitian QQM. Furthermore, non-anti-hermitian QQM has been proven to have a well-defined
classical limit.

The existence of either sources or sinks of 
density of probability has been ascertained to be responsible for the breakdown of the Ehrenfest theorem, and these sources
of probability density are generated by the imaginary terms of the scalar potential
of the Hamiltonian operator.

There are many directions for future research. The results indicate that meaningful quantum quaternionic effects can be researched in physical situations
that are found in non-hermitian CQM, like resonances and scattering phenomena \cite{Bender:2007nj,Moiseyev:2011nhq}.
Another important possible source of interesting physics problems involves geometric phases, and an initial theoretical
 study has already been conducted \cite{Giardino:2016xap}. 
The relation of the present results with the various proposals for non-hermitian CQM 
\cite{Scolarici:2003wu,Scolarici:2009zz,Solombrino:2002vk,Mostafazadeh:2001jk,Mathur:2010nhh,Mathur:2014rnh} and 
non-anti-hermitan QQM \cite{Blasi:2005alt,Blasi:2004qah} must be ascertained as well.
Rigorous mathematical studies are also important. The study of quaternionic
multi-particle states have already been conducted \cite{Giardino:2012ti}, and the spectral theorem has also been considered for
quaternionic operators \cite{Colombo:2016qst}. Nevertheless, a rigorous study of the hermiticity of operators whose
expectation value is given by  (\ref{A8}) is also highly desirable, and are accordingly important directions for
future research.
Measurable effects have never been researched for non-hermitian quaternionic physical situations, 
and we expect that the framework we propose may be useful for renewing the interest in QQM within the field of experimental physics.
On the other hand, we hope that theoretical interest in quaternionic quantum solutions may also be renewed,
particularly the search for explicit solutions.


%
%
%
%

\bibliographystyle{unsrt} 
\bibliography{bib_nahqqm}

\end{document}